# Slow dynamics in a single glass bead

John Y. Yoritomo and Richard L. Weaver
Department of Physics
University of Illinois, Urbana, IL 61801

**Abstract**

Slow dynamic nonlinearity is ubiquitous amongst brittle materials, such as rocks and concrete, with cracked microstructures. A defining feature of the behavior is the logarithmic-in-time recovery of stiffness after a mechanical conditioning. Materials observed to exhibit slow dynamics are sufficiently different in microstructure, chemical composition, and scale (ranging from the laboratory to the seismological) to suggest some kind of universality. There lacks a consensus theoretical understanding of the universality in general and the log(time) recovery in particular. Seminal studies were focused on sandstones and other natural rocks, but in recent years other experimental venues have been introduced with which to inform theory. One such system is unconsolidated glass bead packs. However, bead packs still contain many contact points. The force distribution amongst the contacts is unknown. Here, we present slow dynamics measurements on a yet simpler system—a single glass bead confined between two large glass plates. The system is designed with a view towards rapid control of the contact zone environment. Ultrasonic waves are used as a probe of the system, and changes are assessed with coda wave interferometry. Three different methods of low-frequency conditioning are applied; all lead to slow dynamic recoveries. Results imply that force chains do not play an essential role in granular media slow dynamics, as they are absent in our system.

## I. Introduction

About twenty-five years ago, it was reported that rocks demonstrate a fascinating diminished stiffness after a mechanically-induced conditioning ("pumping"), followed by a gradual log(time) recovery [1,2]. These behaviors are known as slow dynamic nonlinearity and appear to be not unique to rocks. Slow dynamics has been observed in many other materials that possess an internal geometry composed of many contact points, such as concrete, cement and cracked glass. In spite of the ubiquity of these behaviors, they are poorly understood.

The original studies on non-classical nonlinear elasticity in rocks and cement-based materials were performed by the geophysics group at Los Alamos National Lab (LANL) [2–9]. They employed Nonlinear Resonant Ultrasound Spectroscopy (NRUS), which enabled tracking of the test samples' fundamental vibration frequencies (at a few kHz). Application of minor conditioning strain (as little as $10^{-6}$) led to a drop in elastic modulus as revealed by a decrease in the resonant frequency. In much of the LANL work, the conditioning strain was created by oscillatory vibrations, also at a few kHz. More significantly, the drop in modulus was followed, after the strain was removed, by a slow recovery towards the original value. The healing occurred over periods ranging from a few seconds to hours after the conditioning strain was removed, and progressed with the logarithm of time since the conditioning ended. The same behaviors have been observed after conditioning by temperature and humidity changes [10]. Neither the diminished stiffness, nor the recovery, nor its time dependence are well understood.

Other laboratory techniques besides NRUS have been used to monitor changes and recoveries. Lobkis and Weaver [11] monitored slow dynamic recoveries of narrow-band ultrasonic Larsen frequency in sandstone and cement paste after impact conditioning. Using the Larsen frequency, i.e., the ringing in an ultrasonic feedback circuit, enabled ultra-fine time resolution monitoring. Log(t)-like recovery was detected as soon as three milliseconds after the impact. Tremblay et al. [12], after conditioning by impacts in concrete, monitored broad-band diffuse reverberant ultrasonic signals and measured changes using coda wave interferometry. Shokouhi et al. [8] used Dynamic Acousto-Elasticity Testing (DAET) [13,14] in which changes and recoveries were monitored by measuring the transit time of a high-frequency ultrasonic pulse, and fit the observed relaxations to a discrete set of exponential relaxations.

Seismic (at ~1 Hz) wave speed near a fault after an earthquake exhibited a similar loss of stiffness and log(t) recovery, e.g. refs. [15,16]. The recoveries were monitored over periods from days to years and correlated with aftershocks. This behavior is not well understood either.

Slow dynamics (and more generally the unusual nonlinearity of rocks) is believed to arise from the glassy contacts between crystallites and the breaking of bonds due to the pump conditioning, like that seen in dry friction [17]. But beyond that there is little consensus, particularly in regard to the actual physical mechanism of the recovery or the nature of the bonds. TenCate et al. [3], noted that the slowness might be due to a distribution of activation energies associated with atomic-scale barriers that are overcome by thermal fluctuations. This model implies that the log(t) relaxation should proceed at a rate proportional to temperature, though their attempts to measure such dependence were inconclusive. Others have proposed similar models [18–21]. All require the activation energy distribution to be uniform over some range. For example, if the recovery should be logarithmic from 1ms to 1 year, the distribution should be flat from 0.5 to 1.1 eV. Neither a physical justification for such uniformity nor why it should be so universal has been given.

That slow dynamics is time-scale free and robust against significant changes to system details (e.g., microstructure, chemical composition, length scale) could suggest the presence of interacting healing sites and a corresponding criticality [22–24], in which healing rate diminishes in time with a universal exponent $t^{-1}$. One could posit that slow dynamics arises from crackling noise—discrete events that span a broad range of sizes. A diverse number of physical systems are known to exhibit crackling noise, including earthquakes, magnets, and crumpled paper [23,24]. One could further speculate that the 1/t Omori law for aftershock rate [25]—derived as a critical phenomenon [26]—would translate as a log(t) dependence for degree of healing, if the aftershocks are measures of healing rate.

Other models have been proposed and various environmental factors have been suggested as relevant for determining the physical mechanism [27–32]. Of particular note is moisture. Bittner [29] showed that fully saturated cements did not exhibit slow dynamics and has suggested that diffusion of water vapor along cracks is responsible for the slowness. However, other studies of an almost fully-dried sandstone sample—held in vacuum for months—did not show loss of slow dynamics, suggesting water is not responsible [4].

There remains little consensus on the micro-physics ultimately responsible for slow dynamics' remarkable log(t) linearity and ubiquity—in spite of the many phenomenological fits and some plausible hypotheses. This is due in part to limited understanding of the microstructures involved. Rocks and cements are highly complex multi-phase materials, in general consisting of water, crystallites, cracks, inclusions, glassy contacts, residual stresses, and slow chemical reactions.

It is therefore worthwhile to introduce simpler structures for the study of slow dynamics. In particular, unconsolidated glass bead assemblages have been proposed. The structure and internal contacts of bead packs are better understood than the crack geometries of the rocks, cements, and bulk glasses. Depending on pore size the packs may also allow ready and controlled ingress of heat and water vapor. Slow dynamics has been observed in such systems [32–36]. Johnson and Jia [33], using NRUS at 17kHz, were the first to present evidence of slow dynamics in bead packs. Their recoveries, while highly irregular, appeared logarithmic from minutes to hours. Unfortunately, the measurements to date have tended to be irregular and hard to reproduce cleanly. Glass bead packs are further complicated by their complex albeit fascinating acoustics [36,37]. Nonlinearity is strong, especially at low static confining pressures. Even the linear regime is complex; high frequency waves are strongly scattered and highly diffuse [38–40]. We recently reported a new study of slow dynamics in glass bead packs, using experimental methods similar to those presented below (i.e. Sec. III), that has greater precision and lower noise than the studies referenced above [41].

However, can an even simpler structure be used to study slow dynamics? Though we understand the character of the contacts within bead packs better than those in rocks and concrete, there are still many such contacts. Here we propose a new system to study slow dynamics that has only two contact points. It is comprised of a single glass bead confined between two glass plates (Fig. 1). The system is probed using low-amplitude ultrasonic waves, and changes within the system are assessed using coda wave interferometry. This new system is simple enough to suggest a model for the ultrasonics (see appendix). We believe the system may permit rapid control of the environment at the contact points. It will also provide an indication of whether force chains play a role in slow dynamics. Force chains arise in granular systems, [42,43] but should not arise in this two-contact system. If slow dynamics is observed in the system, it would suggest that force chain rearrangements do not play an essential role.

This paper presents studies of slow dynamic recoveries in this simplified, two-contact system. Slow dynamics is observed for three different methods of conditioning. We suggest this system could be a useful venue for examining slow dynamic dependence on various structural and environmental parameters, thereby informing theory for its microphysical basis. In the next section we describe the experimental design, and the resonant transmission of linear ultrasound through the glass bead. The following section presents the coda wave interferometry technique [44–46] for measuring changes in the transmitted diffuse ultrasonic waveform. In section IV we present the results of the slow dynamics measurements. We conclude with a discussion of the advantages of this system for slow dynamic investigations and some implications.

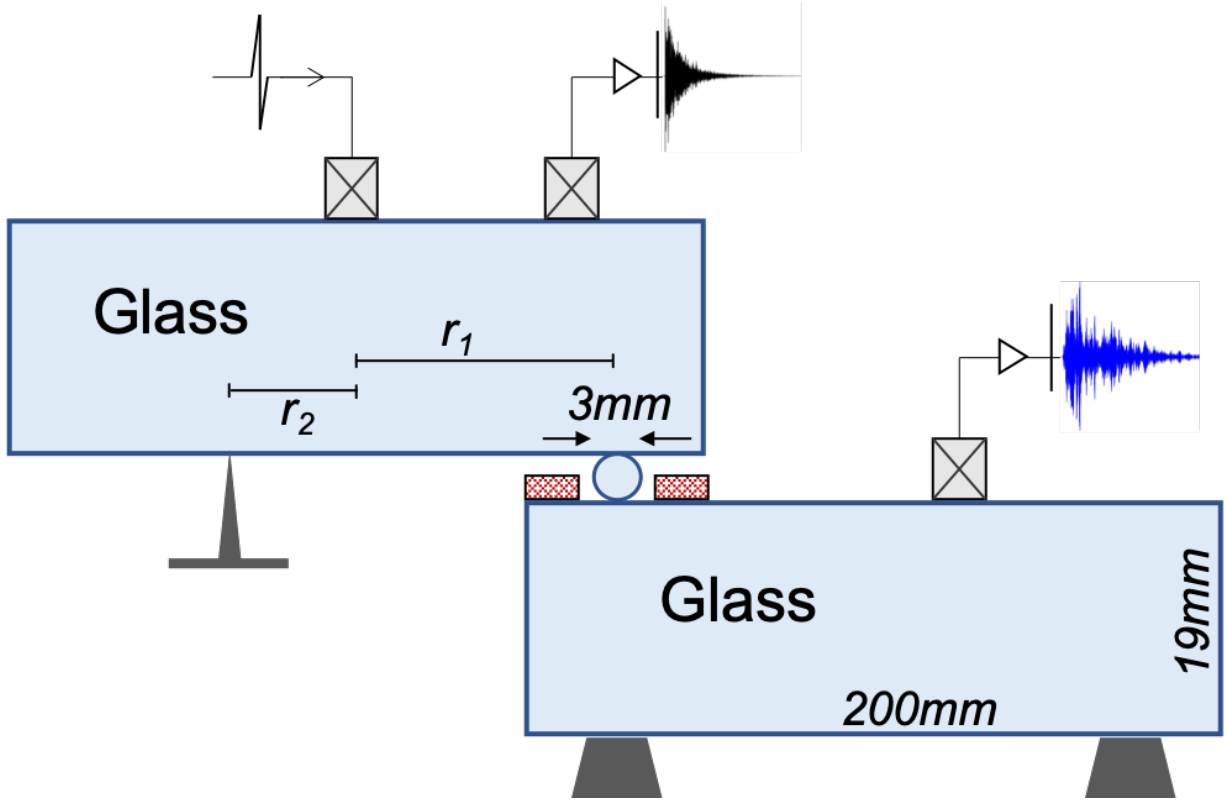

Figure 1. A schematic of the experimental setup (not to scale). A glass plate rests on top of two thumbtacks (one directly behind the other in the picture) and a single glass bead, which in turn rests on a top of an identical glass plate. The lower plate rests on small rubber feet. The plates have dimensions 19mm x 200mm x 155mm. The bead has a diameter of 3mm. A 1.6mm-thick rubber sheet (red and cross-hatched in Fig. 1) is placed between the two slabs in the region where they overlap. A hole is cut in the rubber sheet so that the bead directly touches the glass plates and does not touch the rubber. Two ultrasonic transducers are placed on the upper slab, and one transducer is placed on the lower slab. A 10ns-duration high-voltage broadband ultrasonic pulse is sent to the source transducer every 0.1 seconds. The received signals in the upper and lower slabs are amplified by 40dB preamplifiers and then digitized. The signal in the lower plate depends on transmission through the bead and is markedly different from the signal in the upper plate. The horizontal distances from the bead and the thumbtacks to the upper plate's center of mass are labeled $r_1$ and $r_2$, respectively.

## II. Experimental Setup and Diffuse Ultrasonic Transmission

Two glass plates are coupled by a single glass bead (see Fig. 1). The left side of the top plate rests on two thumbtacks, and the bottom plate rests on small pieces of hard rubber. Each plate is 19mm x 200mm x 155mm and has a mass of 1.5kg. The glass bead has a diameter of 2.97mm and a mass of 30mg. Both the plates and the bead are made of common soda-lime glass. From the locations of where the thumbtacks and bead support the upper plate, the weight from the upper slab on the bead is estimated F = 3.4N.[1] To reduce transmission from the upper slab through the air to the lower slab, a piece of 1.6mm thick rubber is placed between the slabs. A hole is cut out of the rubber sheet so that the bead makes direct contact with both slabs (and not with the rubber). The entire system is placed on a vibration isolation table, as we found that the ultrasound transmission through the bead is sensitive to ambient vibrations from the laboratory floor.

---

[1] This force is obtained by force and torque balance. The bead is approximately in the center of the out-of-plane direction of Fig. 1. Thus, one-dimensional force balance gives $F + f = M g$, where $F$ is the force of the bead on the plate, $f$ is the force of the thumbtacks on the plate, $M$ is the mass of the plate, and $g$ is standard gravity. Torque balance gives $r_1 F = r_2 f$, where $r_1$ ($r_2$) is the distance of the bead (thumbtacks) from the plate's center of mass—see Fig. 1. Thus, $F = Mg/(1 + r_1/r_2) = 3.4\ N$, with $M = 1.5\ kg$, $r_1 = 6.75 cm$, and $r_2 = 2 cm$.

Using our estimate for the contact force of 3.4N, Hertzian contact theory provides an estimate for the radius of the contact circle, $a = \left(\frac{3FR(1-\nu^2)}{2E}\right)^{1/3}$, where R is the radius of the glass bead, E is the elastic modulus of glass, and ν is the Poisson ratio of glass [47]. Using $E = 70$ GPa, $\nu = 0.22$, $R = 1.5$mm, $F = 3.4$N gives a contact radius of $a = 47$ μm. A maximum pressure ($p_0 = 3F/(2\pi a^2)$), the maximum shear stress ($= 0.31 p_0$), and the maximum tensile stress ($= (1/3)(1-2\nu)p_0$) can also be estimated. We find these to be 735MPa, 228 MPa, and 137 MPa, respectively. The maximum shear stress would occur at the center of the contact circle a distance $0.48a = 23$ μm below the center of contact. The maximum tensile stress occurs at the edge of the contact circle. The maximum tension is above the nominal tensile strength of glass (tens of megapascals). Thus, we should expect to see ring cracks in the glass plates and glass bead. However, we do not observe any cracks when we inspect the plates and bead under a microscope after disassembly. Either our glass is stronger than 137MPa, or the cracks close after disassembly and become invisible.

It is an open question as to whether the slow dynamics observed here (Sec. IV) is due to microcracking in the bead and/or slabs, or is a function of contact mechanisms at the bead-slab interface. For future work, we suggest the use of tempered glass, at least at the contact points, as tempered glass is less likely to crack at these loads.

The source ultrasonic transducer is placed on the top glass plate. A receiver transducer is placed on the top plate, and another on the bottom plate. All transducers are coupled with oil. The source and receiver on the bottom are Physical Acoustics Corp. (Mistras) micro30 with a quoted best response from 150 to 400kHz. The receiver on the top plate is a Digital Wave B-1025 with a quoted range of 50 – 2000kHz. A 10ns-duration high-voltage broadband ultrasonic pulse is sent to the source transducer every 0.1 seconds. The received signals are amplified by 40dB ultrasonic preamplifiers and then recorded by a digitizer at 10Msamples/sec. We repetition average 20 received signals to improve signal-to-noise. A repetition-averaged ultrasonic signal is produced approximately every 3 seconds (one second is consumed by the acquisition software). Changes in the signals recorded at the bottom slab are quantified using coda wave interferometry (next section); these changes are used to assess how the transmission responds to conditioning (section IV).

Figure 2 shows a typical ultrasonic signal received on the top plate (Fig. 2 inset) and through the bead on the bottom plate. The signals extends out to 15ms or more. The two signals have a different character: the energy on the top side arrives immediately and decays while the transmitted energy rises over the first two milliseconds and then decays. The signal on the bottom has a more irregular envelope, suggesting it is more narrow-band.

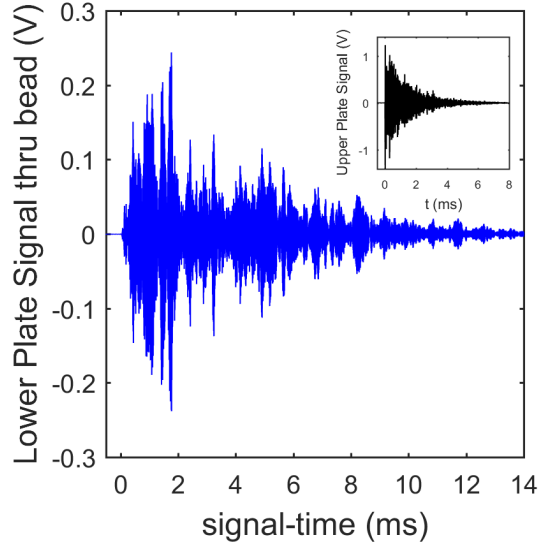

Figure 2. Typical ultrasonic signal that is received at the lower plate. The signal has been delayed and filtered by the bead which couples the upper and lower plate. The inset shows the signal received in the upper plate. The energy on the top arrives practically instantaneously and decays while the transmitted energy rises over the first two milliseconds and then decays.

We expect the bead to significantly modify the ultrasound that is transmitted to the bottom glass plate. Transmission should only occur near the resonances of the bead. These resonances are predicted by Hertzian contact theory, which enables calculation of the rigid body modes of the bead. There should be three modes: the first corresponding to vertical motion perpendicular to the plane of the plates, a second (doubly degenerate) mode corresponding to horizontal motion parallel to the plane, and a third (also doubly degenerate) mode corresponding to rotational motion around horizontal axes. The respective frequency of each mode is given by

$$f_\perp = \frac{1}{2\pi}\sqrt{\frac{2k_\perp}{m_b}} \qquad f_\parallel = \frac{1}{2\pi}\sqrt{\frac{2k_\parallel}{m_b}} \qquad f_{\rm rot} = \frac{1}{2\pi}\sqrt{\frac{2k_\parallel}{2m_b/5}} \qquad (1)$$

where $k_\perp$ ($k_\parallel$) is the stiffness perpendicular (parallel) to the plane of the glass plates and $m_b$ is the mass of the bead. The factor 2/5 in the denominator of $f_{\rm rot}$ arises from the moment of inertia of a solid sphere. The total stiffness is $2k$ (rather than just $k$) because stiffness is present at both the upper and lower contact points.

Hertzian contact theory provides formulas for contact of a solid sphere and a half space. For the case in which the sphere and half space are made from the same material, the force-displacement relation is

$$F = \frac{2}{3}\frac{E}{(1-v^2)}R^{1/2}\delta^{3/2} \qquad (2)$$

where $E, \nu, R$ are defined above, and $\delta$ is the mutual approach of distant points. $F$ is the vertical contact force. Thus, the vertical (perpendicular) stiffness is given by

$$k_\perp = \frac{\partial F}{\partial \delta} = \frac{3F}{2\delta} = \left(\frac{3RE^2}{2(1-\nu^2)^2}\right)^{1/3} F^{1/3} \tag{3}$$

The horizontal (parallel) stiffness is related to the perpendicular one by [48]

$$k_\parallel = \frac{2(1-\nu)}{(2-\nu)} k_\perp \tag{4}$$

The resonant frequencies (eqn. (1)) of the three modes scale with the sixth root of the force: $f \sim F^{1/6}$.

Taking the previously estimated contact force $F$ to be 3.4 N and using $E = 70\ GPa$, $\nu = 0.22$, $R = 1.5mm$, we find that $\delta = 1.47\ \mu m$ and $k_\perp = 3.5 \times 10^6 N/m$. Thus, $f_\perp = 76kHz$, $f_\parallel = 72kHz$, and $f_{\text{rot}} = 113kHz$.

We expect the spectrum of the signal through the bead to have three narrow frequency bands centered at the frequencies of the three rigid body modes of the bead. Figure 3 shows the observed lower slab spectrum (blue curve) as well as the spectrum of the signal received in the upper plate (black curve) for comparison. The wide-band spectrum of the upper plate reflects the wide-band nature of both the transducers and the propagation in the upper plate. The frequency content in the lower plate is very different. As expected, the spectrum is narrow band. However, we identify only one peak near 110kHz—presumably corresponding to the rotational mode of the bead—rather than three (see inset in Fig. 3). The absence of the lower modes is curious. It is possibly due to the transducer's weak response at 70kHz. In support of this hypothesis we increase the force on the bead, by placing an extra 2kg on the upper plate (such that $F$ is now 23N) and then identify three peaks near the (new) expected frequencies (see Fig. 4). The first internal resonance of the bead [49] is also distinguishable near 900kHz (not shown), though transmission there is very weak. Our measurements reported below (Sec. IV) were all conducted at the 3.4N load with a new bead and different position between plates.

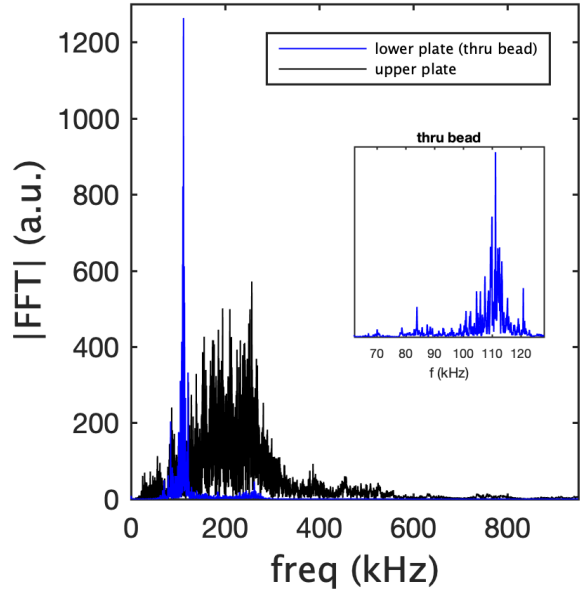

Figure 3. The spectrum of a typical signal received in the lower plate (blue curve) and one in the upper plate (black curve). The wide-band spectrum of the upper plate reflects the wide-band nature of the transducers. The narrow-band spectrum of the lower plate indicates that the bead only allows transmission in a narrow frequency range. The inset expands the spectrum in the lower plate around the frequency region near the resonant frequencies predicted by eqn. (1): $f_\perp = 76 kHz$, $f_\parallel = 72 kHz$, and $f_{rot} = 113 kHz$. Only a single peak, near 110kHz, can be identified—presumably corresponding to the rotational mode.

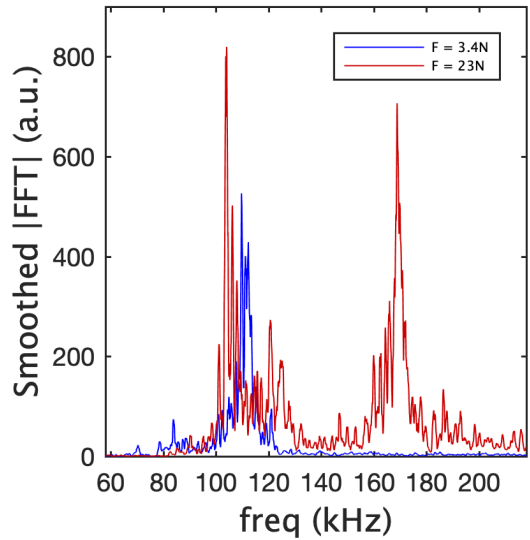

Figure 4. Comparison of the spectrum received at the lower plate when the only weight on the bead is the upper plate (blue curve) and when an additional 2kg is added (red curve). The additional weight increases the force from 3.4N to 23N. Three peaks can be discerned when extra weight is added. Each one is identified with a rigid-body mode of the single bead, and predicted by Hertzian theory to be $f_\perp = 105 kHz$, $f_\parallel = 98 kHz$, and $f_{rot} = 156 kHz$.

## III. Coda Wave Interferometry

Coda wave interferometry (CWI) is used to quantify changes in this two-contact-point system. Figure 5 summarizes the process. A normalized cross-correlation $X_n^i$ is constructed between a reference signal $\phi$ and all signals produced in an experiment $\psi_n$ ($n = 1, 2, ...$) captured at laboratory-times $T_n$. The cross-correlation is calculated over a certain signal-time window $i$ centered at $t^i$ and having a width $W$:

$$X_n^i(\tau) = \frac{1}{A_n^i} \int_{t^i - W/2}^{t^i + W/2} dt\, \phi(t)\psi_n(t + \tau) \tag{5}$$

where A is the normalization factor:

$$A_n^i = \sqrt{\int dt\, \phi^2(t) \int dt\, \psi_n^2(t + \tau)} \tag{6}$$

The integrals in A are over the same region in $t$ as those in the numerator of $X_n^i$. We distinguish between "laboratory-time" and "signal-time" to emphasize the different time scales involved. Laboratory-time $T_n$ ranges from seconds to minutes and its index $n$ goes from 1 to N, where N is the total number of signals captured in a measurement. N is typically between 100 and 400. (Laboratory-time is labeled as "time" in Figs. 6-8, below.) Window times $t^i$ range from 100s of microseconds to milliseconds, and the index $i$ labels a window of signal-time. The range of $i$ varies with window width $W$ and how much signal is being examined. $t$ is signal-time after the main bang of the pulser and ranges from 0 to 14ms. For the results shown below $W = 200$ μs, and the first signal-time window begins at $t^1 - W/2 = 50$μs. We specify the range of $i$ below.

We take the reference signal $\phi$ to be the first signal in an experiment, always before conditioning is applied. Figure 5a plots an example comparison between the reference signal $\phi$ (blue curve) and the 60[th] signal $\psi_{60}$ (red curve), recorded about 180 seconds after $\phi$. Small differences exist between the signals. The plot is shaded to highlight the windows ($W = 200$ μsec) used to construct $X_{60}^{i=1,2,...}$. Figure 5b is an expanded region of panel (a) to show that $\psi_{60}$ is delayed with respect to $\phi$. The delay is quantified by calculating the lapse-time value where $X_n^i$ is maximum, $\mathcal{T}_n^i = \arg\max X_n^i(\tau)$. Figure 5c shows $X_{60}^5(\tau)$ with $\mathcal{T}_{60}^5$ designated by the red vertical line. The $\mathcal{T}_n^i$ values for a given $n$ are plotted versus signal-time $t^i$ (Fig. 5d, for $n = 60$). We fit the ordered pairs $(t^1, \mathcal{T}_n^1), (t^2, \mathcal{T}_n^2)$ ... to a straight line:

$$\mathcal{T}_n^i = \sigma_n t^i + b_n \tag{7}$$

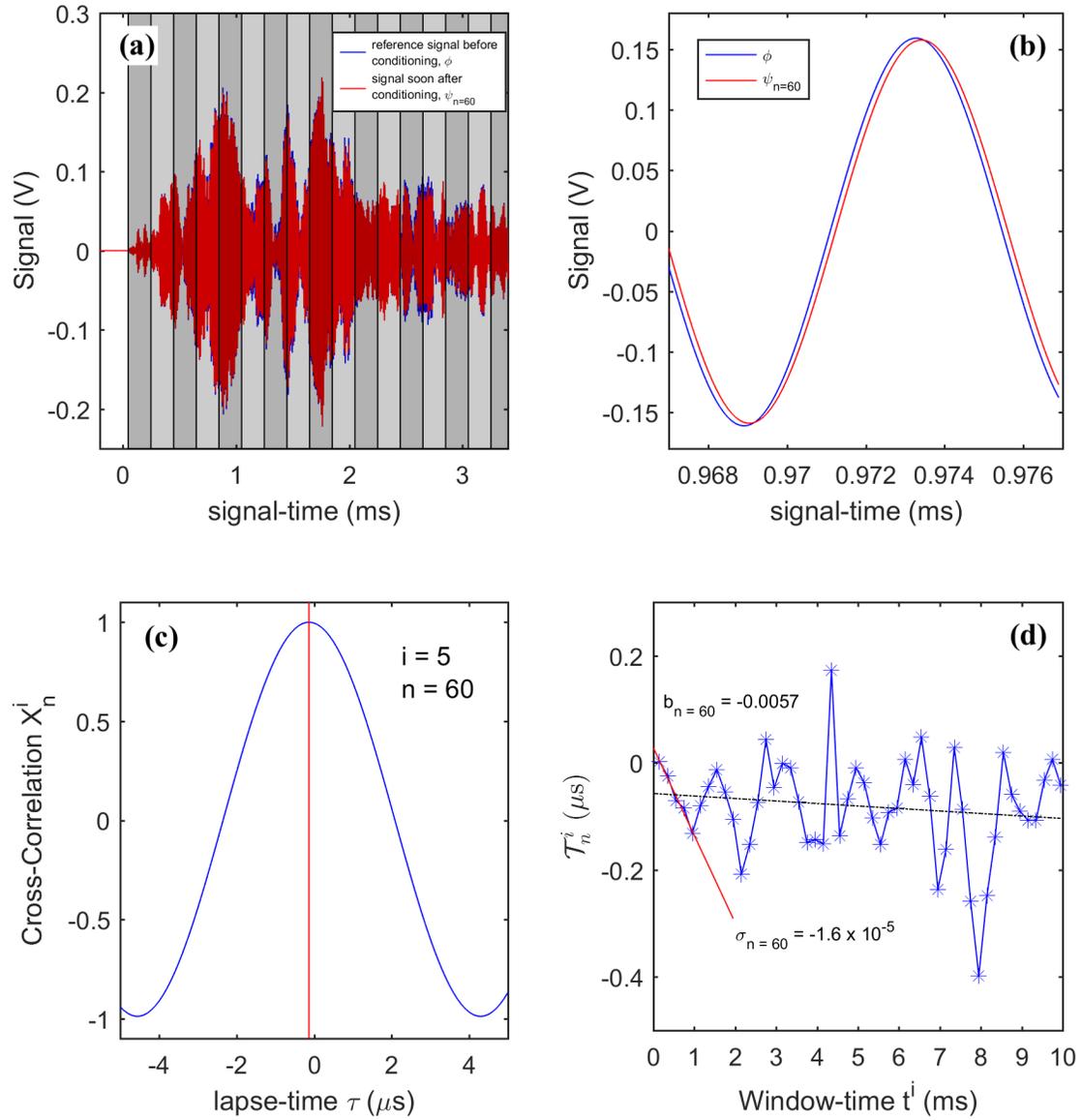

Figure 5. Summary of the coda wave interferometry procedure. Panel (a) shows the reference signal $\phi$ (blue curve) and the 60$^{th}$ signal $\psi_{60}$(red curve) in a measurement. The plot is shaded to highlight the windows ($W = 200\ \mu sec$) used to construct $X_{202}^{i=1,2,\cdots}$. Panel (b) shows an expanded region of panel (a) to demonstrate that $\psi_{60}$ is delayed with respect to $\phi$. The delay is quantified by calculating the lapse-time value where $X_n^i$ is maximum, $\mathcal{T}_n^i = \arg\max X_n^i(\tau)$. Panel (c) shows $X_{60}^5(\tau)$ with its maximum at $\mathcal{T}_{60}^5$ designated by the red vertical line. Thus, in the 5$^{th}$ signal time window, the 60$^{th}$ capture $\psi_{60}(t)$ is delayed relative to $\phi(t)$ by about 0.1 $\mu sec$. The $\mathcal{T}_n^i$ values for a given $n$ are plotted versus window-time $t^i$, as shown in the blue curve of panel (d) for $n = 60$. We fit the early time part of the curve (i.e., signal-time up to 1ms) to a straight line (solid red curve): $\mathcal{T}_n^i = \sigma_n t^i + b_n'$. The slope $\sigma_n$ may be called early time stretch. It is one signature of changes in the system. (The y-intercept $b_n'$ is close to zero and not meaningful.) As shown in Appendix, this stretch is not equal to the true bead stretch. We also fit the entire curve, i.e. signal-time up to 10ms, to an additional straight line (dashed black curve): $\mathcal{T}_n^i = \sigma_n' t^i + b_n$. The y-intercept $b_n$ is a second signature of changes in the system. (Now the slope $\sigma_n'$ is not meaningful.) Thus, we have two signatures of slow dynamics: short-time stretch $\sigma_n$ and y-intercept $b_n$.

For most studies that employ CWI, the plot of $\mathcal{T}_n^i$ versus $t^i$ fits well to a straight line with $b_n = 0$, e.g., [41,44,50]. The slope is then called waveform dilation or "stretch." It can be interpreted as a relative change in wave speed and therefore a relative change in modulus: $\sigma = \frac{\Delta v}{v} = \frac{1}{2}\frac{\Delta M}{M}$. The plot is linear when conditioning affects every part of the system so a wave is continuously delayed during its propagation. However, this is not the case for our system. Here the only components affected by conditioning are the single bead and its contact points with the plates. Thus, after a wave has passed through the bead it is no longer dilated; subsequent propagation within the lower plate is unchanged. Similarly, propagation within the upper plate before transmission is not dilated either. Consequently, at late times, i.e., when the CWI analysis is extended out to 10ms or so in signal-time, we expect not a stretch but rather a shift. This average shift may be identified with the typical ray's sojourn in the bead times the intrinsic stretch within the bead. What we observe, however, is a highly erratic $\mathcal{T}_n^i$ versus $t^i$ with a net average shift, quantified by the y-intercept value b. We also observe a persistent slope σ at early times (t < 1ms) with zero y-intercept. For these early times, the recorded signal is dominated by waves that have spent a significant fraction of their propagation in the bead. This stretch is not equal to the true bead stretch, though presumably related to it. For more discussion, we refer the reader to the Appendix.

An early time stretch, followed by an overall shift, is observed in the plots of $\mathcal{T}_n^i$ versus $t^i$ (see Fig. 5d). We fit the early time part of the curve to a straight line (solid red curve in Fig. 5d): $\mathcal{T}_n^i = \sigma_n t^i + b_n'$, where $\sigma_n$ may be defined as short-time stretch. (The y-intercept $b_n'$ is close to zero.) We also fit the entire curve to a second straight line (dashed black curve in Fig. 5d): $\mathcal{T}_n^i - \mathcal{T}_n^1 = \sigma_n' t^i + b_n$, where now the relevant parameter is the y-intercept $b_n$. We subtract $\mathcal{T}_n^1$ from $\mathcal{T}_n^i$ to compensate for additional noise that is due to uncontrolled trigger jitter associated with the data acquisition. Forcing the first point in the plots of $\mathcal{T}_n^i$ versus $t^i$ equal to be zero is a reasonable constraint, as there should be very little difference between signals for the first 200μs.

Thus, we have two signatures of waveform differences: short-time apparent stretch and the y-intercept. As will be shown in the next section, both quantities, short-time stretch $\sigma_n$ and y-intercept $b_n$, exhibit slow dynamics. The only parameter that differs when calculating stretch or shift is the range of i in signal-time, $t^i$. For stretch, i goes from 1 to 5, where $t^5 = 0.95$ms; for shift, i goes from 1 to 50, where $t^{50} = 9.95$ms.

## IV. Slow Dynamics Results

To examine if our plate-bead-plate system exhibits slow dynamic nonlinearity, we probe with the ultrasonic waves described above. As confirmed *aposteriori*, the ultrasound is of sufficiently low amplitude to ensure that the probe waves themselves are not significantly conditioning the bead.[2] Three types of conditioning, or "pumps," are employed: impulsive, harmonic, and quasi-static (see Table 1). The first two were chosen to correspond to pump methods used by others [4,9,11,33].

---

[2] We confirmed that the ultrasound was not itself conditioning the single bead system by beginning an experiment with the pulse amplitude low. After a sufficient number of repetition-averaged pulses to establish a consistent value of stretch, the pulse amplitude was approximately quadrupled. The stretch values after the pulse amplitude was quadrupled did not show any changes. We took this to be sufficient evidence that the ultrasound was not itself conditioning the single bead system.

The same three pump types were used in a recent study of slow dynamics in unconsolidated glass bead packs [41].

Estimates for the strain produced from each conditioning method are given below. We define conditioning strain under a small change $\Delta F$ (in Newtons) in contact force as

$$\epsilon = \frac{\Delta \delta}{R} = \frac{\Delta F}{k_\perp R} = 1.9 \times 10^{-4} \Delta F \tag{8}$$

This is the *gross* strain across the bead (as opposed to the microscopic strains[3]) and can be compared with strain estimates that have been reported in the literature, such as [3].

| Type of conditioning | Description | $\epsilon$ – Estimated strain | $m$ – Slope of early-time stretch vs. ln(T) plot | Slope of y-intercept vs. ln(T) plot |
|---|---|---|---|---|
| Impulsive | Dropped a 6.35mm diameter wooden ball from 20cm | $\epsilon_{peak} = 3.3 \times 10^{-5}$ | $2.5 \times 10^{-5}$ | $0.0083 \mu s$ |
| Harmonic | Shaker placed on bottom plate | $\epsilon_{rms} = 8.4 \times 10^{-6}$ | $5.7 \times 10^{-5}$ (shaker off) | $0.025 \mu s$ *(shaker off)* |
| Quasi-static | Added and subtracted 65g mass | $\epsilon = 1.2 \times 10^{-4}$ | $1.7 \times 10^{-4}$ (adding weight) $1.3 \times 10^{-4}$ (subtracting weight) | $0.052 \mu s$ *(adding weight)* $0.051 \mu s$ *(subtracting weight)* |

Table 1. Summary of the different conditionings used in the slow dynamics measurements (Sec. IV). The strains associated with each conditioning and the slopes of the recovery are given. The slopes were estimated by fitting the recovery from 15 seconds after recovery to 3 minutes after recovery for impulsive and harmonic conditioning. For quasi-static the slopes were estimated by fitting the recovery from 15 seconds after recovery to 2.5 minutes after recovery. The method for estimating the strains for each conditioning are described in the subsections of Sec. IV.

---

[3] The microscopic strains, e.g., $\epsilon_{micro} = \Delta p_0 / E$, are somewhat greater but vary from place to place within the contact region and scale differently with $F$ and $R$.

## A. Slow dynamics from impulsive conditioning

Our impulsive pump is a wooden ball (mass of 90mg, diameter of 6.35mm) dropped from 21cm above the upper slab near where the slab is coupled to the glass bead. Impulsive pumping has been previously used on cement paste and sandstone by dropping similarly sized wooden balls [11] and on concrete by dropping small metal balls [12]. Primary benefits of impulsive pumping are a clear time of conditioning and ease of application [11,12].

Results for the impulsive conditioning are shown in Fig. 6. Slow dynamics is observed, as the characteristic drop in stretch followed by a slow recovery is clearly present in Fig. 6a. Stretch measurements have a rms deviation from a smoothed fit of about $5 \times 10^{-7}$, which we interpret as the precision. The recovery is also clearly logarithmic in time (Fig. 6b): $\sigma = m \ln(T_n - T_z) +$ constant, where the slope $m = 2.5 \times 10^{-5}$ and $T_z$ was chosen to give good linearity at early times.[4] The slope was estimated by fitting the recovery from 15 seconds after impact to 3 minutes after impact. The time for full recovery, i.e. when the curve in Fig. 6b would cross the σ=0 axis, can be estimated as 4 hours. Observation of full recovery is, however, difficult due to potential contamination by drifts in temperature and/or residual recoveries due to earlier mechanical disturbances. We do not attempt it here. The results for y-intercept $b$, rather than stretch, are shown in Fig. 6c and 6d. The same behavior is observed.

The strain induced by the ball drop is estimated by placing an accelerometer directly over the bead during a ball drop. The accelerometer records a signal $a(t)$. A second accelerometer was placed over the thumbtacks (see Fig. 1). The acceleration recorded there was much smaller than $a(t)$, indicating that the position of the thumbtacks can be taken as the pivot point $O$. If the bead is a horizontal distance $l$ from $O$, the torque around $O$ is $\Delta F\, l = I_O a(t)/l$. The moment of inertia is $I_O = I_{cm} + M(r_2)^2$, so $\Delta F(t) = (0.731 kg) \times a(t)$, where $L = 0.2m$ is the length of the slab, $r_2 = 0.02m$, $l = 0.0875m$ and $M = 1.5 kg$. The force signal looks approximately like a single cycle sine wave with a peak value of $\Delta F_{peak} = 0.174N$. Using equation (8), we obtain a peak strain of $\epsilon_{peak} = 3.3 \times 10^{-5}$.

---

[4] We do not have independent measure of the zero time $T_z$ (the time of impact, the time at which harmonic conditioning starts or ceases (Sec. IV.B), or the time at which the quasi-static loads are changed (Sec. IV.C)). The data itself, however, clearly indicate this zero time to within the three second interval between data points. In the plots (panel b and d of Figs. 6-8 and Fig. 9) we adjust $T_z$ so as to make them fully linear. Different choices for $T_z$ within the known interval will distort the linearity only for the first few data points. Strictly speaking, we only demonstrate linearity for $T_n - T_z > \sim 15s$. It is recommended that future work record the zero time independently, e.g., [11]. A precise record of the zero time can be one of the advantages of impulsive conditioning.

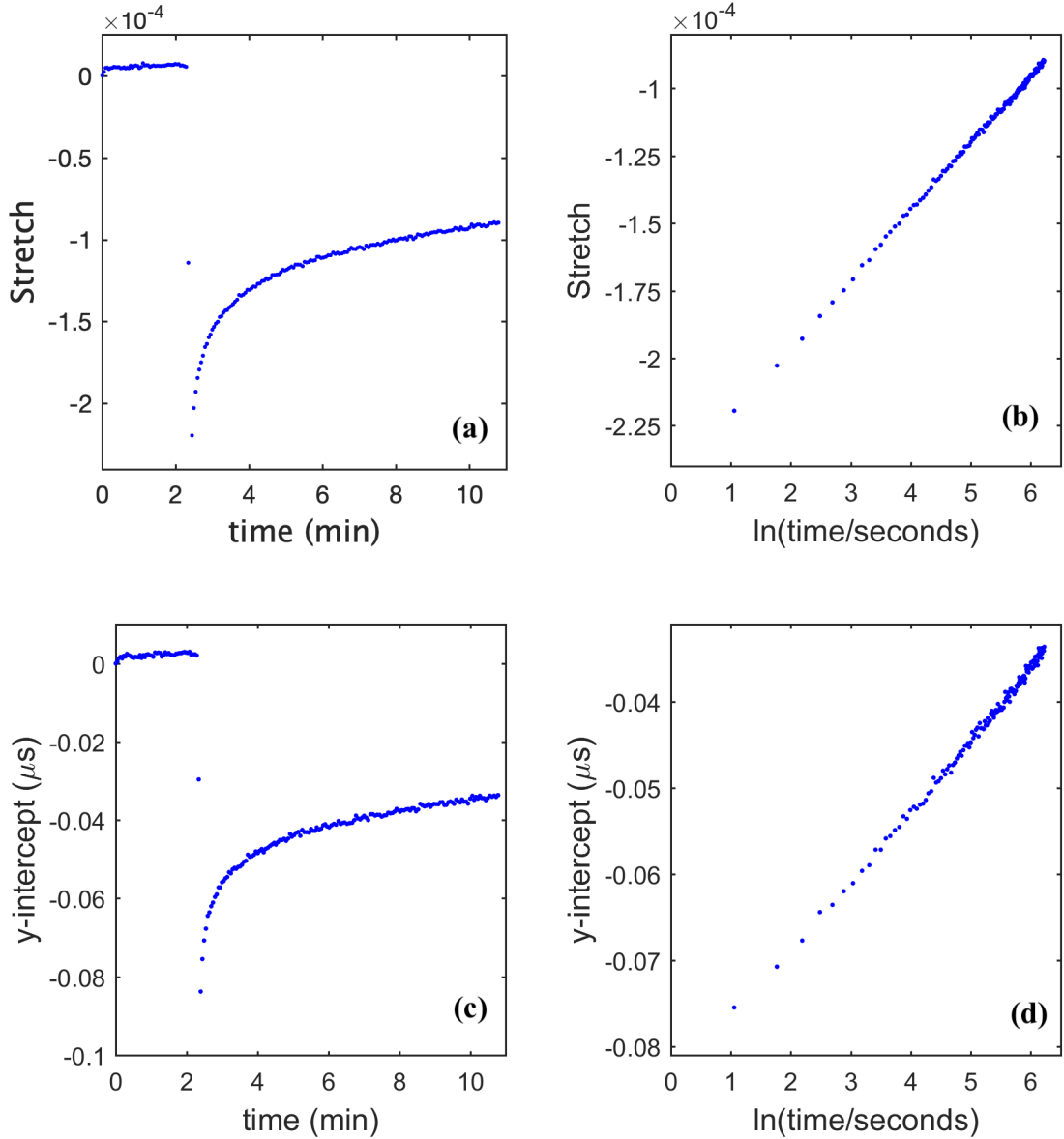

Figure 6. The slow dynamics results for impulsive pumping. Stretch (over the first millisecond), $\sigma_n$, is plotted versus laboratory-time, $T_n$, in panel (a). The recovery is logarithmic in time, as seen in panel (b): stretch versus the logarithm of time after the ball drop at $T_{drop} \sim 2.5$ minutes. Stretch measurements have precision of about $5 \times 10^{-7}$. Panel (c) and (d) show the slow dynamic results using the y-intercept, $b_n$, instead of stretch. Recovery in intercept is also $\ln(T)$. The slopes (Table 1) were estimated by fitting the recovery from 15 seconds after recovery to 3 minutes after recovery.

**B. Slow dynamics from harmonic conditioning**

Harmonic pumping is provided by a shaker that rests on the lower slab and vertically vibrates a 166g mass at 200Hz. Much of the work from LANL [2–5,9,33] used harmonic pumping, as it can be integrated with NRUS. In those experiments a sustained sinusoidal excitation at the longitudinal resonance of the sample was used for conditioning. One advantage of harmonic pumping is the ability to control and easily measure how much pump strain the sample is experiencing.

The results for our harmonic conditioning are shown in Fig. 7. The shaded regions in Fig. 7a and 7c indicate when the shaker was on. Again, slow dynamics is observed in the stretch recoveries while the shaker was off. Figure 7b shows that the stretch recovery is logarithmic in time with a slope of $m = 5.7 \times 10^{-5}$ for the last recovery. As in impulsive conditioning, the slopes were estimated by fitting the recovery from 15 seconds after recovery to 3 minutes after recovery. Slow dynamics is also observed in y-intercept measurements (Fig. 7c and 7d). An extrapolated time for full recovery with harmonic conditioning is difficult to determine. It is not clear to what value stretch is recovering, as it was for impulsive conditioning: zero value of stretch. For harmonic conditioning, the quiescent state is distorted by previous cycles of conditioning and relaxation. However, some cycling is necessary, as the sample must first reach a steady state; one period of conditioning is not sufficient (Fig. 7a).

As with the impulsive conditioning, the strain induced by the shaker can be estimated by attaching an accelerometer, $a(t)$, to the upper slab over the bead contact point and a second one over the thumbtacks. The acceleration recorded over the thumbtacks was again much smaller than $a(t)$, so the point $O$ is still a pivot point. Using the same reasoning as previously, $\Delta F(t) = (0.731 kg) \times a(t)$. We estimate the rms force—from the observed rms $a(t)$—as $\Delta F_{rms} = 0.0441 N$. The rms strain is thus $\epsilon_{rms} = 8.4 \times 10^{-6}$.

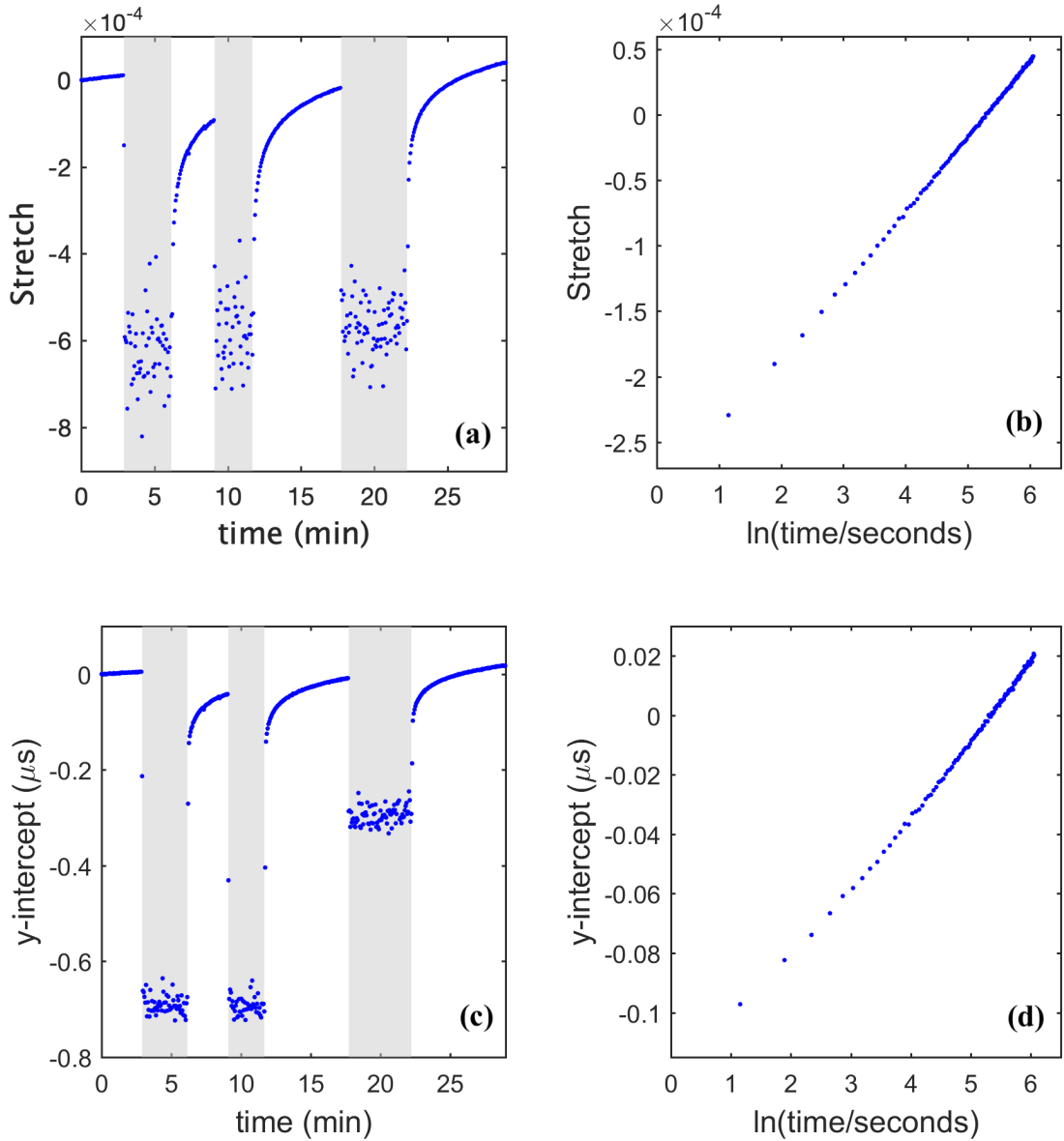

Figure 7. The slow dynamic results for harmonic pumping. Stretch (over the first millisecond), $\sigma_n$, is plotted versus laboratory-time, $T_n$, in panel (a). The shaded regions indicate conditioning, i.e., when the shaker was on. The recovery is logarithmic in time, as seen in panel (b): stretch versus the logarithm of time for the last recovery section. Precision in stretch when the shaker is off is better than $10^{-6}$. Panel (c) and (d) show the slow dynamic results using the y-intercept, $b_n$, instead of stretch. Recovery in intercept is also log(t). The slopes (Table 1) were estimated by fitting the recovery from 15 seconds after recovery to 3 minutes after recovery.

**C. Slow dynamics from quasi-static conditioning**

The quasi-static pump involved the periodic placing and removing of an additional 65g mass on the top plate directly over the bead. The results are shown in Fig. 8. The shaded regions in Fig. 8a and 8c indicate when the 65g was added. It is expected that relative wave speed would increase when the extra mass is added (frequency $f \propto F^{1/6}$, according to Hertzian theory [37]) leading to a positive stretch. However, that $f$ would continue to increase logarithmically (Fig. 8b) after the addition signifies nontrivial slow dynamics. Similarly, when subtracting the mass, we expect stretch to return to its initial value, i.e. zero. Rather, the measured value overshoots the expected value and then recovers slowly towards it. Similar, albeit noisier, behavior is observed in shift (Fig. 8c and 8d). Slope estimates are given in Table 1. They were estimated by fitting the recovery from 15 seconds after adding or subtracting the additional mass to 2.5 minutes after adding or subtracting. The strain for quasi-static conditioning can be estimated directly from $\Delta F = (0.065 \text{kg}) \times (9.81 \text{m/s}^2) = 0.64 N$, giving $\epsilon = 1.2 \times 10^{-4}$.

For quasi-static conditioning we can also predict how much stretch should occur long after adding the additional mass. By Hertzian theory, frequency $f$ should be proportional to the sixth root of the force. The addition of 65g increases the static force on bead approximately 19%, so the fully relaxed stretch should be 0.031. If we take the short time dilation σ of the transmitted wave to be one quarter (see Appendix) the actual bead stretch $\delta f/f$, and assume $\delta f/f$ goes to a value of 0.031 eventually, then an extrapolation of the observed log(time) relaxation rate $m = 1.7 \times 10^{-4}$ from σ = 0.0046 at log(time) = 3 to σ = 0.00775 requires 70 years. This striking number, and its great difference from the 5hr extrapolated time to full recovery for impulsive conditioning (Sec. IVA), begs to be explained. We observed the same peculiarity after quasi-static conditioning in the glass bead pack for which an extrapolated time to full recovery had a similar, conspicuously large, magnitude [41]. This extrapolation is, however, quite sensitive to the poorly understood difference between apparent stretch σ and the actual bead stretch $\delta f/f$.

The results, especially Fig. 8a, show clearly the symmetry breaking of the inducing source, as both tensile and compressive forces lead to a relaxation characterized by a slow dynamic *increase* in modulus regardless of the sign of the pumping. This asymmetry has been emphasized by Tencate *et al.* [3] as a key characteristic of slow dynamics and distinguishes it from other creep phenomena. We observe the same symmetry breaking in an unconsolidated glass bead pack under quasi-static conditioning [41].

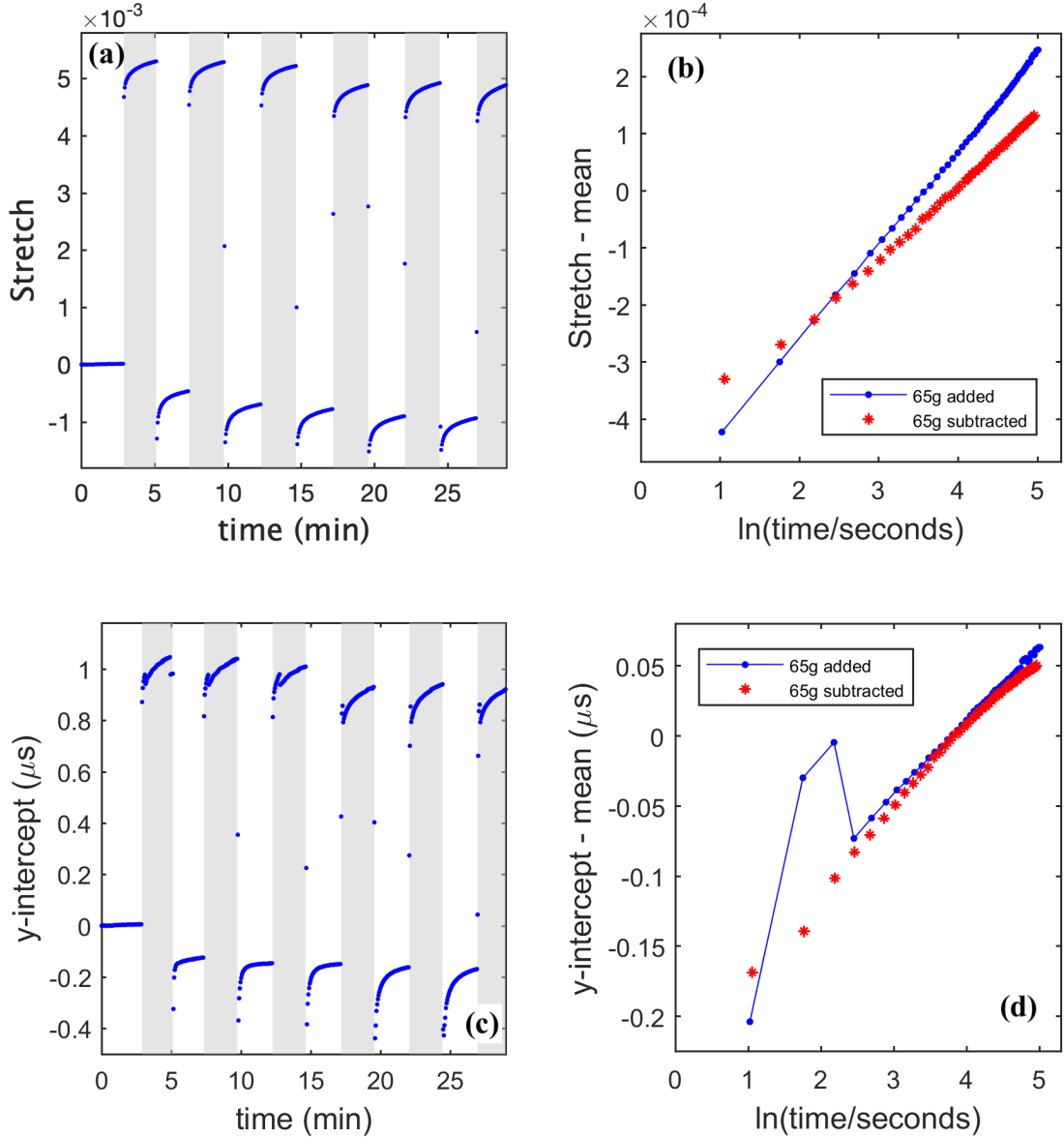

Figure 8. The slow dynamic results for quasi-static pumping. Stretch (over the first millisecond), $\sigma_n$, is plotted versus laboratory-time, $T_n$, in panel (a). The shaded regions indicate the times in which the 65g mass was placed on the top plate. Both conditioning (65g added) and recovery (65g subtracted) are logarithmic in time. Panel (b) shows stretch minus its mean versus the logarithm of time for the last conditioning and recovery sections. The mean is subtracted to more easily plot both curves in one panel. Like the other conditioning measurements (Figs. 6 and 7), stretch has a precision greater than $10^{-6}$. Panel (c) and (d) show the slow dynamic results using the y-intercept, $b_n$, instead of stretch. Recovery in shift is also log(t), though with some peculiar and unexplained irregularity for the first few seconds after conditioning by adding weight (panel (d), solid line). The slopes (Table 1) were estimated by fitting the recovery from 15 seconds after recovery to 2.5 minutes after recovery.

## V. Discussion

It is important to confirm that the slow dynamic relaxation was occurring at the contact points between the bead and the plates, rather than elsewhere in the system. When we examine the signals received on top (signals like that shown in the inset in Fig. 2) using CWI, no slow dynamics is observed in the upper plate.[5] This indicates that neither the glass plate itself nor the thumbtacks supporting it exhibit slow dynamics. Moreover, when using the same pump-probe scheme (with any of the three pumps) on only a single glass plate, we do not observe any slow dynamics. This eliminates the possibility that slow dynamics was occurring between the lower plate and the rubber feet supporting it.

In a recent study of slow dynamics in a glass bead pack system [41], we expounded three advantages of that system over the more commonly used materials to study slow dynamics (natural rocks, cements, concrete). Those advantages were i) a simplified chemistry and history, ii) many internal surfaces that are easier to characterize, and iii) a high porosity and large pores, enabling better control of the environment at the contact points. The single bead system presented here offers those same advantages to an even greater degree. As with glass bead packs, the chemistry is straightforward and there is virtually no history—the system is formed once the top plate rests on the single bead. The contacts can be modeled with Hertzian contact theory, and because there are only two contacts, it is possible to formulate a model for the ultrasonic changes (see Appendix). It can be difficult to control, or even know, the internal structure within cement paste or sandstones, as the conditions internally may be different from those on the surface. These difficulties do not arise in our single bead system, as air can easily enter the contact area and equilibrate with its surroundings. The system is also small enough to enclose easily for controlling its environment.

To demonstrate the usefulness of this experimental venue, we conclude with a preliminary investigation into the role of fluid at the bead-plate contacts. This was motivated by the work of Bittner [29] and Bittner and Popovics [30], who imaged the transport of moisture within small (1-10mm) cement paste discs before and after harmonic conditioning. They observed the migration of water out of micron-size pores during conditioning and the slow migration back into the pores afterwards. Our investigation examines the effects of hydrophobic surface treatments on slow dynamic recovery. If water at the contacts is a key part of the mechanism, one expects that a hydrophobic surface would behave differently from a normal one. Our results are shown in Fig. 9. A normal glass bead was replaced with a glass bead having a hydrophobic coating. Three different hydrophobic coatings were used: Silcotek's Dursan, Silcotek's experimental coating, and Surfactis's Seesurf. We also placed a bead coated in oil between the plates. The same qualitative slow dynamic behavior was observed regardless of the coating. That slow dynamics remains after surface treatments perhaps indicates that it is microcracking that is responsible for the behavior. However, as mentioned above, we did not find any microcracks when we inspected the plates and bead under a microscope after disassembly. There remains the possibility that cracks formed under static load and then closed before we inspected the bead or plates. More careful monitoring for cracks is indicated for future work.

---

[5] At times before the Heisenberg time in the upper plate, ~12ms, most rays in the upper plate have not visited the bead. So any slow dynamics in that wave field due to the bead would be diluted there. Moreover, only frequencies near the bead resonances are significantly exposed to the bead's properties, and that would further dilute the effect of the bead's slow dynamics.

To our knowledge, the single bead system possesses the lowest number of contact points to exhibit slow dynamics. Previous granular systems, unconsolidated or consolidated, that exhibited slow dynamics had many contact points. One could conjecture that slow dynamics in bead packs could be an indication of changing force chains [42,43]. Our results demonstrate that force chains cannot be an essential mechanism behind slow dynamics. No force chains could arise in this simplified system, yet slow dynamics remains. However, we should emphasize that although there are only two *macroscopic* contact points, presumably—depending on roughness—each macroscopic contact point consists of many microscopic ones. Zaitsev et al. [32] reported many nanoscale asperities on a single glass bead. We further recall the reader's attention to the possibility that the contact areas have been cracked by the tensile stresses at their perimeter and that the slow dynamics is a process occurring at these cracks.

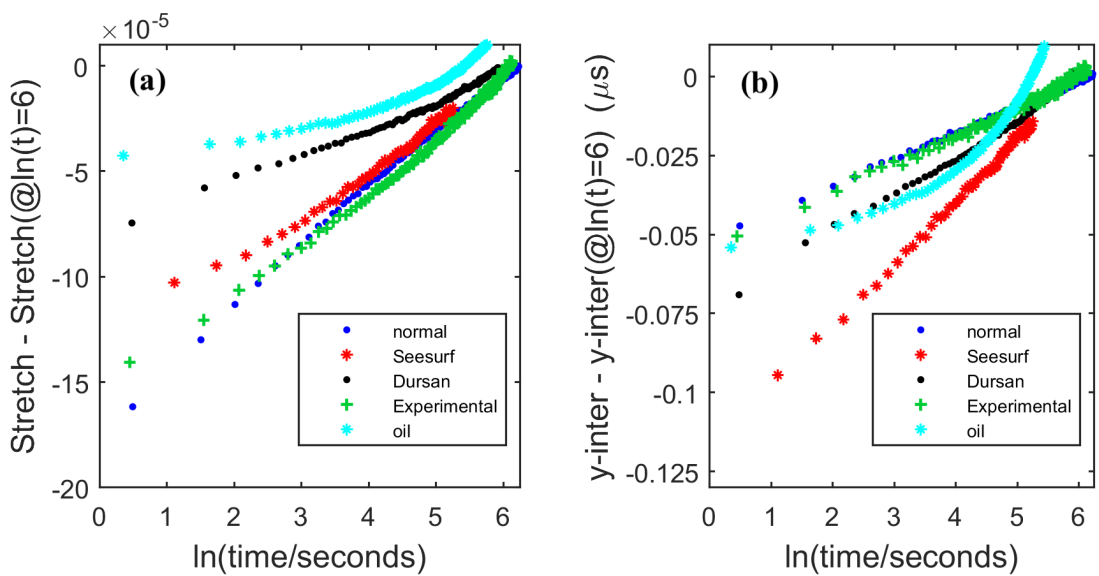

Figure 9. The slow dynamics recovery results after impulsive conditioning for different kinds of bead coating. Both stretch (panel (a)) and y-intercept (panel (b)) are plotted versus the log(time). Both are also normalized so that all curves reach zero at $\ln(T) = 6$. Three different hydrophobic coatings were used: Surfactis' Seesurf, Silcotek's Dursan, and Silcotek's experimental coating. A bead without any coating (labeled "normal) is plotted for comparison. Results for a fifth bead coated in oil are also shown. For the oiled bead the recovery is not linear after 20 seconds or so. This may be due to the oil flowing during the measurement. Different bead coatings do not qualitatively affect the log(time) recovery.

**VI. Conclusion**

It has been demonstrated that a single bead system exhibits the nonlinear elastic phenomena of slow dynamics. The use of ultrasonic wave probes with CWI processing provides low noise and great sensitivity to changes (of the order $10^{-6}$ or less in stretch) in the bead. Slow dynamic response has been observed after a variety of low frequency pumps—not just harmonic but also impulsive and quasi-static pumping. We believe that this venue has a number of benefits. It is particularly suited for determining slow dynamic dependence on environmental parameters, like humidity and temperature. Our results indicate that force chains do not play an essential role in the physical mechanism behind slow dynamic recovery.

## Acknowledgements

We are most grateful to John Popovics and James Bittner for their interest and encouragement and for many fruitful discussions.

## Appendix: Theory and numerics for conditioned transmission through a bead between two plates

We suggest a model for the diffuse wave transmission through a single glass bead between two slabs, and how that transmission is modified by property changes (conditioning and relaxation) at the bead and its contacts. Numerical simulations are found to be in accord with laboratory measurements that show the time delay of the signal in the lower slab after conditioning is linear in signal age, at least until 1ms, and then becomes erratic.

We express the signal in the lower slab as a convolution of three time series. A propagation $G^U$ in the upper slab from the source to the bead, a resonant transmission T through the bead, and a propagation $G^L$ from the bead to the receiver in the lower slab. The signals before and after conditioning are (at times such that few rays have interacted more than once with the bead) given by

$$\phi(t) = G^U(t) \otimes T(t) \otimes G^L(t) \qquad (A1a)$$

$$\psi(t) = G^U(t) \otimes T'(t) \otimes G^L(t) \qquad (A1b)$$

$T$ and $T'$ are the transfer functions through the bead before and after conditioning. We write them in the frequency domain as $T(\omega) = [(\omega - i\gamma)^2 - \omega_d^2]^{-1}$ and $T'^{(\omega)} = [(\omega - i\gamma')^2 - \omega'^2_d]^{-1}$, where $\omega_d$ and $\omega'_d$ are the bead's damped natural frequencies before and after conditioning, and $\gamma$ and $\gamma'$ are the corresponding resonance widths. At times before the Heisenberg time[6] rays are unlikely to visit the same site, so $G^U$ and $G^L$ are well represented as centered white noises $\eta$ and $\nu$ under exponentially decaying envelopes representing absorption:

$$G^U(t) = \eta(t)\exp(-\sigma^U t) \qquad (A2a)$$

$$G^L(t) = \nu(t)\exp(-\sigma^L t) \qquad (A2b)$$

The above model assumes transmission through a single resonant bead mode. However, the bead has at least five nontrivial rigid body modes (Sec. II). Measured spectra (Fig. 3) suggest the chief transmission is at the rotational resonance, of which there are two modes. In order to model transmission via two modes, we would modify the above:

$$\phi(t) = G^{U1}(t) \otimes T_1(t) \otimes G^{L1}(t) + G^{U2}(t) \otimes T_2(t) \otimes G^{L2}(t) \qquad (A3)$$

---

[6] The Heisenberg time is equal to the modal density $2\pi dN/d\omega$, where $N$ is the mode count. Heisenberg time is about 12.3ms at 110 kHz for glass slabs of volume $589 cm^3$ and surface area $755 cm^2$.

with a similar formula, though with $T_1'$ and $T_2'$, for the signal $\psi(t)$ in the conditioned structure. Here $G^1$ and $G^2$ differ in having statistically independent white noises $\eta$ and $\nu$ corresponding to the two horizontal polarizations of the elastic waves at the bead: $G^{U1}(t) = \eta_1(t)\exp(-\sigma^U t)$, $G^{U2}(t) = \eta_2(t)\exp(-\sigma^U t)$, $G^{L1} = \nu_1(t)\exp(-\sigma^L t)$, and $G^{L2}(t) = \nu_2(t)\exp(-\sigma^L t)$. $T_1$ and $T_2$ are as indicated above, with closely spaced or identical resonant frequencies and widths.

It is a straightforward matter to evaluate the above model numerically. We take the absorptions $\sigma$ to have the independently measured values: $\sigma^U = 345/s$ and $\sigma^L = 472/s$. Waveforms $\phi$ and $\psi$ are constructed with a time spacing $dt = 0.1$ μs using random number generators for the white noises. Convolutions are performed by multiplying in the frequency domain.

Figure A1 shows a typical waveform and its spectrum (|FFT|) as constructed using two modes, with close—but not fully degenerate—resonant frequencies $\omega_{d1} = 628000/s$ and $\omega_{d2} = 1.02\omega_{d1}$. We choose the bead loss rates of each resonance to be equal: $\gamma_1 = \gamma_2 = \gamma = 1256/s$. The figure shows that the spectrum is narrow, in accord with $\gamma \ll \omega_{d1}$, with most of the amplitude confined to the region around $100 kHz$. But under the envelope of those Lorentzians, we see irregular features that may be identified as Erickson noise associated with the longer lived reverberant fields in the slabs. In the time-domain these features manifest in a signal (Fig. A1a) which is a narrow band process at 100 kHz with an irregular envelope, a beat pattern amongst the Erickson peaks and between the two modes.

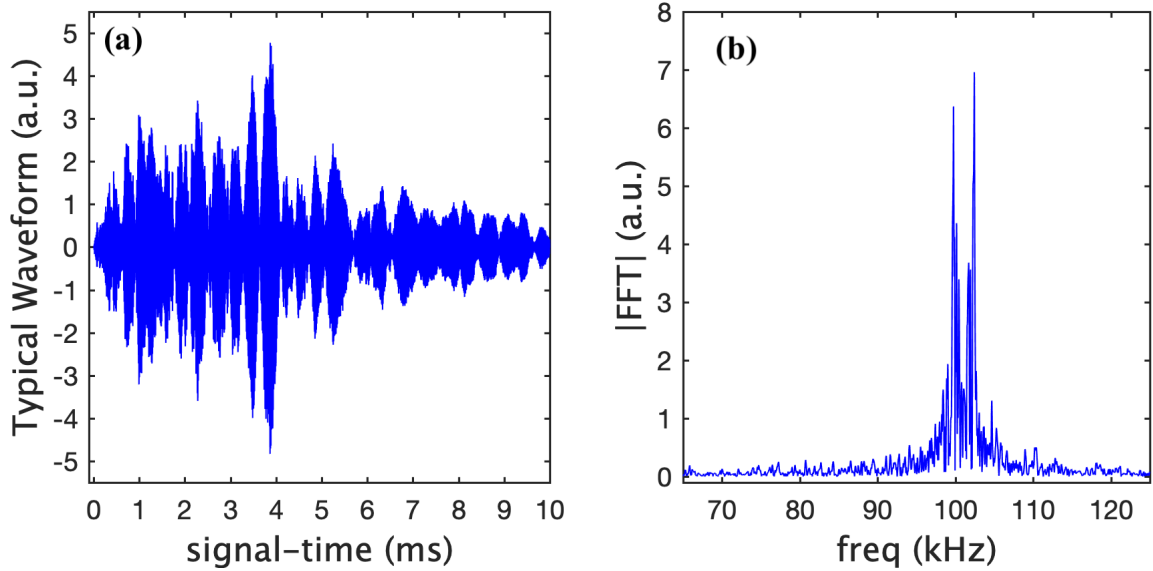

Figure A1. A waveform (a) and its spectrum (b) constructed numerically from eqn. (A3).

Figure A2 shows the delays $\mathcal{T}$ extracted from a pair of signals $\phi$ and $\psi$ for the case of two bead modes (at 100kHz and 102kHz, each with $\gamma = 1256$). Each bead mode frequency is stretched in $\psi$ by an amount $\epsilon = 0.0004$ relative to $\phi$: $\omega_{d1}' = \omega_{d1}(1 - \epsilon)$ and $\omega_{d2}' = \omega_{d2}(1 - \epsilon)$. No stretch is applied to the widths: $\gamma' = \gamma$. The several curves are for different realizations of the random number generator. The delays were calculated as described in Section III, with nonoverlapping $200\ \mu sec$ windows.

The behavior is rather like that seen in the laboratory. For the first millisecond the delays are proportional to signal-time. For greater signal-times, the delays become increasingly erratic.

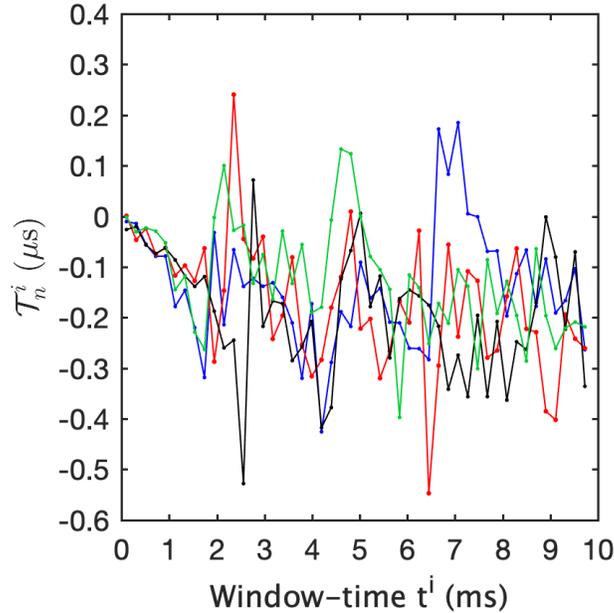

Figure A2. The delays $\mathcal{T}^i$ plotted versus window-time $t^i$. Each curve corresponds to a different realization of the random number generator used to construct η and ν. The delays $\mathcal{T}$, i.e., lapse-time values where $X_n^i$ is maximum, are extracted from a pair of signals $\phi$ (eqn. (A3)) and $\psi$ for the case of two bead modes—with close but not equal resonant frequencies, $\omega_{d1} = 628000 Hz$ and $\omega_{d2} = 1.02\omega_{d1}$, and with equal $\gamma = 1256$. Each bead mode frequency is stretched by an amount $\epsilon = 0.0004$, with no stretch of the widths (such that $\omega'_{d1} = \omega_{d1}(1 - \epsilon)$, $\omega'_{d2} = \omega_{d2}(1 - \epsilon)$ and $\gamma' = \gamma$). The delays were calculated as described in Section III, with nonoverlapping $200\ \mu sec$ windows. The behavior is rather like that seen in the laboratory. For the first millisecond the delays are proportional to signal age. For greater ages, the delays become increasingly erratic functions of time.

The slope of the curves at early time, interpretable as an apparent stretch, are not equal to the bead stretch itself. In the considered example, the slope appears to be about one quarter the bead stretch of 0.0004. Taken at face value one would conclude that the lower slab signal at early times $t$ is composed of waves that have sojourned an average time $t/4$ in the bead, (and the remainder of the time in the slabs where they are not subject to any delays.) That the effective amount of time spent in the bead must be less than 100% is evident; why the fraction should be constant for the first millisecond, or why it should be 25% is less so. Presumably this difference depends on parameters such as σ and γ.

There is much room for further study on this model. What determines the erratic behaviors? What determines the early time non-erratic slope and its duration? What features are sensitive to the presence of two, rather than one, bead mode? Such questions are, however, outside the current scope.

# References

[1]     P. A. Johnson, B. Zinszner, and P. N. J. Rasolofosaon, J. Geophys. Res. B Solid Earth **101**, 11553 (1996).
[2]     J. A. TenCate and T. J. Shankland, Geophys. Res. Lett. **23**, 3019 (1996).
[3]     J. A. TenCate, E. Smith, and R. A. Guyer, Phys. Rev. Lett. **85**, 1020 (2000).
[4]     J. A. TenCate, Pure Appl. Geophys. **168**, 2211 (2011).
[5]     P. Johnson and A. Sutin, J. Acoust. Soc. Am. **117**, 124 (2005).
[6]     R. A. Guyer and P. A. Johnson, Phys. Today **52**, 30 (1999).
[7]     R. A. Guyer and P. A. Johnson, *Nonlinear Mesoscopic Elasticity: The Complex Behaviour of Granular Media Including Rocks and Soil* (2009).
[8]     P. Shokouhi, J. Rivière, R. A. Guyer, and P. A. Johnson, Appl. Phys. Lett. **111**, (2017).
[9]     J. A. TenCate, E. Smith, L. W. Byers, and T. J. Shankland, AIP Conf. Proc. **524**, 303 (2000).
[10]    J. A. TenCate, J. Duran, and T. J. Shankland, in *Nonlinear Acoust. Begin. 21st Century*, edited by O. V. Rudenko and O. A. Sapozhnikov (Moscow State University, 2002), pp. 767–770.
[11]    O. I. Lobkis and R. L. Weaver, J. Acoust. Soc. Am. **125**, 1894 (2009).
[12]    N. Tremblay, E. Larose, and V. Rossetto, J. Acoust. Soc. Am. **127**, 1239 (2010).
[13]    G. Renaud, S. Calĺ, and M. Defontaine, Appl. Phys. Lett. **94**, (2009).
[14]    G. Renaud, P. Y. Le Bas, and P. A. Johnson, J. Geophys. Res. Solid Earth **117**, 1 (2012).
[15]    F. Brenguier, M. Campillo, C. Hadziioannou, N. M. Shapiro, R. M. Nadeau, and E. Larose, Science (80-. ). **321**, 1478 (2008).
[16]    M. Gassenmeier, C. Sens-Schönfelder, T. Eulenfeld, M. Bartsch, P. Victor, F. Tilmann, and M. Korn, Geophys. J. Int. **204**, 1490 (2016).
[17]    L. Bureau, T. Baumberger, and C. Caroli, Eur. Phys. J. E **8**, 331 (2002).
[18]    X. Li, C. Sens-Schönfelder, and R. Snieder, Phys. Rev. B **97**, 1 (2018).
[19]    R. Snieder, C. Sens-Schönfelder, and R. Wu, Geophys. J. Int. **208**, 1 (2017).
[20]    A. V. Lebedev and L. A. Ostrovsky, Acoust. Phys. **60**, 555 (2014).
[21]    L. Ostrovsky, A. Lebedev, J. Riviere, P. Shokouhi, C. Wu, M. A. Stuber Geesey, and P. A. Johnson, J. Geophys. Res. Solid Earth 5003 (2019).
[22]    P. Bak, C. Tang, and K. Wiesenfeld, Phys. Rev. Lett. **59**, 381 (1987).
[23]    J. P. Sethna, K. A. Dahmen, and C. R. Myers, Nature **410**, 242 (2001).
[24]    E. K. H. Salje and K. A. Dahmen, Annu. Rev. Condens. Matter Phys. **5**, 233 (2014).
[25]    B. E. Shaw, Geophys. Res. Lett. **20**, 907 (1993).
[26]    A. P. Mehta, K. A. Dahmen, and Y. Ben-Zion, Phys. Rev. E **73**, 56104 (2006).
[27]    K. E.-A. Van Den Abeele, J. Geophys. Res. **107**, 1 (2002).
[28]    J. Somaratna, Evaluation of Linear and Nonlinear Vibration Methods To, 2014.
[29]    J. A. Bittner, Understanding and Predicting Transient Material Behaviors Associated with Mechanical Resonance in Cementitious Composites, University of Illinois at Urbana-Champaign, 2018.
[30]    J. A. Bittner and J. S. Popovics, Appl. Phys. Lett. **114**, 021901 (2019).
[31]    O. O. Vakhnenko, V. O. Vakhnenko, T. J. Shankland, and J. A. TenCate, AIP Conf. Proc. **838**, 120 (2006).
[32]    V. Y. Zaitsev, V. E. Gusev, V. Tournat, and P. Richard, Phys. Rev. Lett. **112**, 1 (2014).
[33]    P. A. Johnson and X. Jia, Nature **437**, 871 (2005).

[34] X. Jia, T. Brunet, and J. Laurent, Phys. Rev. E - Stat. Nonlinear, Soft Matter Phys. **84**, 2 (2011).
[35] V. Y. Zaitsev, V. E. Nazarov, V. Tournat, V. E. Gusev, and B. Castagnède, Europhys. Lett. **70**, 607 (2005).
[36] V. Tournat and V. E. Gusev, Phys. Rev. E - Stat. Nonlinear, Soft Matter Phys. **80**, 1 (2009).
[37] V. Tournat and V. E. Gusev, Acta Acust. United with Acust. **96**, 208 (2010).
[38] R. L. Weaver and W. Sachse, J. Acoust. Soc. Am. **97**, 2094 (1995).
[39] X. Jia, C. Caroli, and B. Velicky, Phys. Rev. Lett. **82**, 1863 (1999).
[40] X. Jia, Phys. Rev. Lett. **93**, 8 (2004).
[41] J. Y. Yoritomo and R. L. Weaver, ArXiv:1908.08935 (2019).
[42] H. M. Jaeger, S. R. Nagel, and R. P. Behringer, Phys. Today **49**, 32 (1996).
[43] H. M. Jaeger, S. R. Nagel, and R. P. Behringer, Rev. Mod. Phys. **68**, 1259 (1996).
[44] O. I. Lobkis and R. L. Weaver, Phys. Rev. Lett. **90**, 4 (2003).
[45] R. Snieder, Pure Appl. Geophys. **163**, 455 (2006).
[46] T. Planès and E. Larose, Cem. Concr. Res. **53**, 248 (2013).
[47] K. L. Johnson, *Contact Mechanics* (Cambridge University Press, Cambridge, UK, 1985).
[48] A. Merkel, V. Tournat, and V. Gusev, Phys. Rev. E - Stat. Nonlinear, Soft Matter Phys. **82**, 1 (2010).
[49] A. E. H. Love, *Treatise on Mathematical Theory of Elasticity. 4th Edition* (1944).
[50] R. Snieder, A. Grêt, H. Douma, and J. Scales, Science (80-. ). **295**, 2253 (2002).